\documentstyle[preprint,aps,graphicx]{revtex}
\tightenlines
\begin{document}
\newcommand{\bref}[1]{eq.~(\ref{#1})}
\newcommand{\be}{\begin{equation}}
\newcommand{\en}{\end{equation}}
\newcommand{\bs}{$\backslash$}
\newcommand{\us}{$\_$}

\title{Faraday rotation spectra of bismuth-substituted ferrite garnet films with
in-plane magnetization}
\author{L.E. Helseth, R.W. Hansen, E.I. Il'yashenko, M. Baziljevich and T.H. Johansen}
\address{Department of Physics, University of Oslo, P.O. Box 1048 Blindern, N-0316 Oslo, Norway}

\maketitle
\begin{abstract}
Single crystal films of bismuth-substituted ferrite garnets have been synthesized by
the liquid phase epitaxy method where GGG substrates are dipped into the flux. 
The growth parameters are controlled to obtain films with in-plane magnetization
and virtually no domain activity, which makes them excellently suited for 
magnetooptic imaging. The Faraday rotation spectra were measured across the visible
range of wavelengths. To interprete the spectra we present a simple model based 
on the existence of two optical transitions of diamagnetic character, one tetrahedral and one octahedral. 
We find excellent agreement between the model and our experimental results 
for photon energies between 1.77 and 2.53 eV, corresponding to wavelengths
between 700 and 490 nm. It is shown that the Faraday rotation changes 
significantly with the amount of substituted gallium and bismuth. Furthermore, the experimental 
results confirm that the magnetooptic response changes linearly with the 
bismuth substitution.

\end{abstract}

\narrowtext

\newpage

\section{Introduction}
Bismuth-substituted ferrite garnets (Bi:FGs) are well known to have a giant magnetooptical
response\cite{Wittekoek1,Wittekoek2,Takeuchi,Krinchik,Hansen,Hansen1,Hansen2,Hansen3,Simsa}. 
For this reason, they have found widespread use as spatial light modulators, 
optical switchers and optical isolators. Recently it was realized that 
Bi:FG films with in-plane magnetization allow visualization and detection of 
magnetic fields\cite{Belyaeva,Dorosinskii}. As basic sensors for magnetooptic imaging 
they have been used extensively in studies of magnetic flux in superconductors, 
domain formation in magnetic materials, currents in microelectronic circuits 
and recorded patterns in magnetic storage media\cite{Koblischka,Polyanskii,Vlasko-Vlasov,Johansen,Hubert,Egorov,Kotov}. 
In spite of their successful applications, relatively few studies have been 
directed towards the understanding of the magnetooptical properties of 
Bi:FG films themselves. Of particular interest here is the behaviour and
optimization of the Faraday rotation spectra. 

The dodecahedral sites in garnet crystals can accept several rare-earth ions,
and each substitution seems to introduce its own special modification in the 
properties. All of the $Bi^{3+}$, $Ce^{3+}$ and $Pb^{2+}$ have 
been shown to increase the Faraday rotation within certain wavelength 
intervals\cite{Kotov,Simsa0,Gomi,Shinagawa1,Kucera}. With $Bi^{3+}$ a very
strong Faraday rotation is obtained in the visible range, which together with
the fact that the grown crystals are nearly transparent, make them by far the
most frequently used material for magnetooptic imaging. For this application an
important point is also that the films with in-plane magnetization are
essentially free of magnetic domains.

The understanding of the magnetooptical spectra in garnets is based on
the knowledge of the spin orbit interaction in these materials.
The spin orbit interaction generally causes a splitting of both the ground and
excited states. However, in many cases it is possible to assume splitting of 
one of the states. Then only two situations are possible, (i) A paramagnetic transition, which is
due to a orbital degenerate ground state with a singlet excited state. 
(ii) A diamagnetic transition, which is due to a orbital singlet ground state 
with a spin orbit split excited state. Since the paramagnetic and diamagnetic transitions have quite
different lineshapes\cite{Kotov}, it should in principle be possible to 
distinguish between isolated diamagnetic and paramagnetic 
transitions, and therefore obtain a correct interpretation of the spectra. 
However, Bi:FG materials have several transitions within the 
UV and visible region, which complicates the interpretation considerably. This 
has lead to some controversy regarding the kind of transitions involved, and 
therefore how to reproduce the spectra theoretically\cite{Simsa,Dionne1,Dionne,Allen,Matsumoto,Matsumoto1}. 

Simsa $et$ $al.$ measured the Faraday rotation of 
liquid phase epitaxy (LPE) grown Bi:FG films with small substitutions of bismuth\cite{Simsa}. To interprete the 
spectra, they introduced one diamagnetic and several paramagnetic transitions. 
In this way they were able to correctly predict the behaviour of the spectra 
in the visible and near-infrared region.
However, there were still some deviations between experimental and 
theoretical data when the bismuth substitution increased. More recently, 
Matsumoto $et$ $al.$ measured the Faraday rotation for Bi:FG films made by 
the gel method\cite{Matsumoto,Matsumoto1}. Here reasonable 
agreement between the experimental data and the theoretical predictions was
obtained with four paramagnetic transitions in addition to 
the four diamagnetic ones. However, the inclusion of paramagnetic transitions 
was not explained. Actually, Dionne and Allen questioned whether 
paramagnetic transitions can account for the contribution from Bi, since the strong superexchange field of the spin 
degeneracy exclude Zeeman splitting of the ground state, thus leaving a spin 
singlet ground state\cite{Dionne}.   

In this paper we report experiments and modelling of the 
Faraday rotation spectra in Bi:FG films suitable for magnetooptic imaging. To 
that end we have grown a series of Bi:FG films using the LPE technique, and 
characterised their chemical composition. The Faraday rotation spectra have 
been measured for photon energies between $1.77$ and $2.53$ eV, corresponding to
wavelengths between $700$ and $490$ nm. It is shown that the Faraday 
rotation changes significantly with the amount of substituted gallium and 
bismuth. Furthermore, the comparison of experimental and theoretical data 
confirms that the magnetooptic response changes linearly with the bismuth 
substitution.    

\section{Experimental}
\subsection{Sample preparation}
Single crystal films of Bi:FG were grown by isothermal LPE from
$Bi_{2}O_{3}/PbO/B_{2}O_{3}$ flux onto (100) oriented gadolinium gallium
garnet (GGG) substrates. The growth takes place while the 30 mm diameter substrate
is dipped into the melt contained in a Pt crucible. During the growth, the 
parameters could be controlled to create low magnetic coercivity and 
in-plane magnetization in the garnet films. The thickness of the films was
measured using a Scanning Electron Microscope (SEM), and 
their composition determined with an Electron MicroProbe (EMP). Thicknesses and 
compositions of four selected samples are listed in Table 1. For sample 1, 2
and 4 the film was grown on only one side of the substrate, whereas sample 3 has 
a 2 $\mu m$ thick film on both sides of the substrate. 

From Table 1 one sees that the films can be represented by the following 
general formula; 
$\{Lu_{3-x-y-z}Y_{y}Bi_{x}Pb_{z}\}[Fe_{2-u_{a}}Ga_{u_{a}}](Fe_{3-u_{d}}Ga_{u_{d}})O_{12}$,
where $\{ \}$ indicates the dodecahedral site, [ ] the octahedral site, and () the
tetrahedral site. Note that only the total gallium content $u=u_{a}+u_{d}$ can be
extracted from the EMP, and to determine $u_{a}$ and $u_{d}$ separately other 
techniques such as neutron spectroscopy must be applied. Usually, these 
techniques require bulk samples, and are not very useful for thin films.
The films also contain small amounts of $Pb$, typically of the order 
of $z=0.05$. In the theoretical analysis of the Faraday spectra, we will neglect 
the contribution from Pb.  

\subsection{Faraday spectra measurements}
The Faraday rotation as a function of wavelength at room
temperature was measured using a setup consisting of a monochromatic light 
source, polarizer, two magnetic Helmholtz coils, sample, analyzer and detector. A
schematic drawing is shown in Fig. \ref{f1}. The Helmholtz coils produce a 
magnetic field up to 1.1 kOe along the optical axis, which makes it possible 
to saturate the Faraday rotation of the sample. In this work
all the samples were brought into saturation during the measurements. 
The Faraday rotation angle was measured as\cite{Simsa},
\begin{equation}
\Theta_{F}=\frac{\theta_{tot}-\theta_{sub}}{t}\,\,\,\, ,
\end{equation}
where $\theta_{tot}$ is the Faraday rotation measured for the combined film and 
substrate, $\theta_{sub}$ is the Faraday rotation of the bare substrate and $t$ 
is the thickness of the film. The substrates used in this study are  
0.5 mm thick, and give only a small contribution of $\sim 0.1$ degrees. In the 
present study we found that interference effects due to finite 
thickness as well as the Faraday ellipticity could be neglected. The Faraday 
rotation has been measured across the useful visible wavelength range with an 
accuracy of $0.05$ degrees. The wavelength was determined within $\pm 1$ nm. 

Shown in Figs. 2-4 is the observed Faraday rotation as 
a function of energy for samples 1-4. Here $\Theta_{F}$ is denoted by negative
numbers since we follow the sign-convention of refs.\cite{Simsa,Dionne,Allen}. 
One sees that for sample 1-3, the magnitude of 
$\Theta_{F}$ increases monotonically with the photon energy. 
The largest rotation was found in sample 3 at 2.43 eV where $\Theta_{F}=4.6$ 
$deg/\mu m$. Since sample 4 is relatively thin, we were able to measure the
spectrum up to 2.53 eV without reducing the accuracy of the measurement. This revealed a peak 
in the spectrum near 2.45 eV, where $\Theta_{F}=4.1$ $deg/\mu m$.  

\section{Theory}
Our analysis of the Faraday rotation spectra is based on the 
molecular-orbital energy-level diagram of Dionne and Allen\cite{Dionne1}, 
shown in Fig. \ref{f5}. The basic mechanism behind the 
enhanced Faraday rotation is here the cooperative action of $Fe^{3+}$ 
ions with degenerate orbital terms that are split further by covalent
interaction with $Bi^{3+}$. In fact, substitution of bismuth results in an 
overlap between the 6p orbit of $Bi^{3+}$, 2p orbit of $O^{2-}$ and 3d orbit 
of $Fe^{3+}$. As discussed by Shinagawa\cite{Shinagawa}, and later by Dionne 
and Allen\cite{Dionne1}, this overlap results in an spin orbit splitting 
$2\Delta_{i}$ associated with the tetrahedral and octahedral sites, see also 
Fig. \ref{f5}.      
In the model presented here we ignore the contribution from transitions other 
than those introduced by bismuth substitution. In fact, LuIG and YIG have very 
similar Faraday rotation spectra with a positive peak located near 2.8 eV, but 
with very low Faraday rotation below 2.6 eV\cite{Hansen1}. This makes our
assumption reasonable. 

In order to calculate the magnetooptic response, an expression
which links the microscopic parameters to the Faraday 
rotation is needed. In magnetooptics, this is usually expressed through the 
complex permittivity tensor, which for Bi:FG has the form,
\begin{displaymath}
\mathbf{\epsilon} = 
\left( \begin{array}{ccc} 
\epsilon_{0} & \epsilon_{1} & 0 \\
-\epsilon_{1} & \epsilon_{0} & 0 \\
   0       & 0        & \epsilon_{0} \\
\end{array} \right) \,\,\, ,
\end{displaymath}
where $\epsilon_{1}=\epsilon_{1}^{'} +j\epsilon_{1}^{''}$ represents the
magnetooptical response and $\epsilon_{0}$ the non-magnetooptical part.
In the visible region, the Faraday rotation can be written\cite{Dionne},
\begin{equation}
\Theta_{F}\approx \frac{\omega }{2nc} \epsilon_{1}^{'} \,\,\, ,
\label{a1}
\end{equation}
where $\omega$ is the the frequency, or equivalently the energy of the light, c 
is the speed of light in vacuum and $n$ is the refractive index. Independent 
measurements show that it is reasonable to set $n=2.5$ in the visible 
range\cite{Wittekoek2,Hansen1}. 

Assuming that the transitions are of electric dipole nature, the off-diagonal 
elements derived from first order time-dependent perturbation theory are given
by\cite{Dionne},  
\begin{equation} 
\epsilon_{1} =\frac{2\pi Ne^{2}}{m} \sum_{i=a,d} \sum_{+}^{-}
(\pm)f_{i\pm} \frac{\omega
(\omega_{i\pm}^{2}-\omega^{2}-\Gamma_{i}^{2}) +
j\Gamma_{i}(\omega_{i\pm}^{2}+\omega^{2}+\Gamma_{i}^{2})}{\omega_{i\pm}(\omega_{i\pm}^{2}-\omega^{2}+\Gamma_{i}^{2})^{2}
+ 4\omega^{2}\omega_{i\pm}\Gamma_{i}^{2}} \,\,\,.
\label{a2}
\end{equation} 
The inner sum is over the right- and left-hand circular polarized light, 
whereas the outer is over the optical transitions in the 
tetrahedral and octahedral sublattices. The prefactor $(\pm)$ denotes a substraction. 
The $\omega_{i+}$ and $\omega_{i-}$ represent the resonance energy for right 
and left-hand circular polarized light, respectively. Similarly, $f_{i+}$ and $f_{i-}$ are
the respective oscillator strengths, while $\Gamma_{i}$ is the half-linewidth 
of the transition. Furthermore, $e$ and $m$ are the electron charge and mass,
respectively, whereas $N$ is the active ion density.

To first order, it is reasonable to assume that $N$ is directly
proportional to the bismuth content $x$\cite{Dionne}. This assumption has also been 
confirmed by experiments\cite{Hansen1,Hansen2}. Furthermore, it is known that 
the strong enhancement of Faraday rotation is caused by iron-pair transitions, 
involving both octahedral and tetrahedral transitions simultaneously\cite{Scott}. 
Therefore, iron dilution of either sublattice results in a reduction of the
active ion density. For these reasons we propose that the active ion density 
can be written as,
\begin{equation}
N=N_{0}(1-u_{d}/3)(1-u_{a}/2)x  \,\,\, .
\label{a21}
\end{equation}
$N_{0}$ is a constant, and may expected to be 1/3 of the density of rare-earth ions on the 
dodecahedral site, i.e. $1.3\times10^{22}$ $cm^{-3}/3$. When $x=3$, this
interpretation implies that the dodecahedral site is fully occupied by bismuth. 
 
The oscillator strength for right- and left-hand circular polarized light are 
given by\cite{Dionne}, 
\begin{equation}
f_{i\pm}=\frac{m\omega_{i\pm}}{2h}|<g_{i}|\hat{X}|e_{i\pm}>|^{2} \,\,\, ,
\label{a33}
\end{equation}
where $h$ is Planck's constant, $\hat{X}$ is the electric dipole operator, and 
$|g_{i}>$, $|e_{i\pm}>$ being the wavefunction of the ground state and excited 
state, respectively. Using that for the present case one has 
$\omega_{i\pm}=\omega_{i} \pm \Delta_{i}$, we obtain from eq. (\ref{a33}),
\begin{equation}
f_{i\pm}=\frac{f_{i}}{2\omega_{i}}(\omega_{i} \pm \Delta_{i}) \,\,\, ,
\label{a3}
\end{equation}
where $f_{i}$ is an effective oscillator strength\cite{Dionne}. This expression  
is based on neglecting the difference between $|e_{i+}>$ and $|e_{i-}>$.

In order to fully predict the Faraday spectra, one needs detailed
knowledge about the quantum mechanical oscillator strengths. However, this is 
not easily accessible with so many ions involved, and we will instead treat 
them as adjustable parameters in the comparison with experimental data. 
With eq. (\ref{a3}) inserted in eqs. (\ref{a1}) and (\ref{a2}), one obtains
\begin{eqnarray} 
\Theta_{F} &=& \frac{\pi e^{2} \omega ^{2}}{nmc} \sum_{i=a,d}\frac{Nf_{i}}{\omega_{i}} \nonumber \\
& & \times\left\{ 
\frac{(\omega_{i}+\Delta_{i})^{2}-\omega^{2}-\Gamma_{i}^{2}}{\left[(\omega_{i}+\Delta_{i})^{2}-\omega^{2}+\Gamma_{i}^{2}
\right]^{2}
+ 4\omega^{2}\Gamma_{i}^{2}} -
\frac{(\omega_{i}-\Delta_{i})^{2}-\omega^{2}-\Gamma_{i}^{2}}{\left[(\omega_{i}-\Delta_{i})^{2}-\omega^{2}+\Gamma_{i}^{2}\right]^{2}
+ 4\omega^{2}\Gamma_{i}^{2}} \right\} \,\,\,  ,
\label{a4} 
\end{eqnarray}
which represents our model for the Faraday rotation. 
      
\section{Discussion}   
To fit theoretical curves to the experimental data, the product 
$Nf_{i}/x$ was chosen as free parameter. According to eq. (\ref{a21}) 
$N$ depends on the gallium and bismuth content. On the other hand, 
$f_{i}$ is roughly independent on these two substitutions. Therefore, the
product $Nf_{i}/x$ may be adjusted to any substitutions and transitions.
The parameters $\Delta_{i}$, $\omega_{i}$ and $\Gamma_{i}$ were chosen as 
sample independent, and the values suggested by Dionne and 
Allen were used as a starting point in the fitting\cite{Dionne1}. Table 2 
presents the parameters we found to give the best fit between the theoretical 
curves and the experimental data. Note that our values for $\omega_{d}$, 
$\Gamma_{d}$ and $\Gamma_{a}$ differs slightly from those of Dionne and Allen. 
A reason for this is that the minimum Faraday rotation seems to be shifted slightly towards 
lower energies as compared to the data analysed in ref.\cite{Dionne}. 
Furthermore our values for $\Gamma_{d}$ and $\Gamma_{a}$ are smaller to account for the 
narrow linewidth near 2.45 eV seen in Fig. \ref{f4}.  

In Fig. \ref{f2} and \ref{f3} the Faraday rotation calculated from 
eq. (\ref{a4}) is plotted as solid lines, and show an excellent agreement
with experimental data. In Fig. \ref{f2} the dashed and dash-dotted lines
represent the contribution to the total Faraday rotation from the tetrahedral 
and octahedral sites, respectively. Note that below 2.2 eV, the main 
contribution to $\Theta _{F}$ comes from the transition related to the 
octahedral site. 
    
Fig. \ref{f4} shows the theoretical curve (solid line) fitted to the 
experimental data for sample 4. Again there is excellent agreement at small 
photon energies, but some deviation appears above 2.4 eV, where both the 
experimental and theoretical curves show a peaked behaviour. In fact the
experimental data indicate a stronger peak than the theoretical prediction. 
There may be several reasons for such a discrepancy.
First, eq. (\ref{a21}) represents a simplification for the present type of 
FGs. In particular, we have not considered the 
canting of the spins or the temperature dependence of $N$.  
There is also a possibillity that $\Delta_{i}$ and $\Gamma_{i}$ should be 
chosen differently to obtain a stronger peak around 2.45 eV. Finally, it must 
be noted that we have neglected the transitions associated with 
pure LuIG.
 
We have in Fig. \ref{f4} also plotted theoretical predictions for the case 
$x=1.5$ (dashed curve) and $x=2$ (dash-dotted curve) assuming that gallium content is identical
to that of sample 4. When $x=2$ it seems possible to obtain a Faraday rotation 
exceeding 6 $deg/\mu m$ at 2.3 eV in this material.   

The distribution of gallium on the tetrahedral and octahedral
sites have been examined by Czerlinsky and Scott $et$ $al.$ for various 
garnet-compositions with and without bismuth\cite{Czerlinsky,Scott}. Their 
results indicated that gallium will replace $Fe^{3+}$ in both tetrahedral and 
octahedral sublattices, but with the strongest dilution of tetrahedrally coordinated $Fe^{3+}$. In
fact, Czerlinsky found that for $Y_{3}Fe_{4}Ga_{1}O_{12}$ the distribution is
0.9 Ga on tetrahedral sites and only 0.1 on octahedral sites. Although the
samples investigated in this study have somewhat different composition, we will 
assume that the gallium distribution is the same. Then, based 
on eq. (\ref{a21}) the gallium dependence can be removed resulting in a Faraday 
rotation associated only with the Bi-content given by,
\begin{equation}
\Theta_{F}^{Bi} =\frac{\Theta_{F}}{(1-u_{d}/3)(1-u_{a}/2)} \,\,\, ,
\end{equation}
where $u_{d}=0.9u$ and $u_{a}=0.1u$. 
In Fig. \ref{f6} the $\Theta_{F}^{Bi}$ is displayed as a function of the bismuth
substitution. The experimental data have the same notation as in the previous 
figures, and are shown for two different energies, 1.97 eV (630nm) and 
2.3 eV (540nm), corresponding also to the theoretical curves drawn as solid 
and dashed lines, respectively. To obtain the theoretical curves we used  
$N_{0}f_{a}=2.77 \times 10^{23}$ $cm^{-3}$ and 
$N_{0}f_{d}=6.30 \times 10^{22}$ $cm^{-3}$, which are the average values of 
the four samples as calculated from Table 2 and the assumed gallium distribution.    
The experimental points are located very close to the straight lines, thus
indicating that our linear approximation is valid. However, the linear 
approximation may not hold for large x, since saturation phenomena could
change the behaviour. In fact, experimental 
studies by Matsumoto $et$ $al.$ indicated breakdown of the linear 
approximation when $x>2$\cite{Matsumoto,Matsumoto1}. 

\section{Conclusion}
We have synthesized single crystals of bismuth-substituted ferrite garnets by
the liquid phase epitaxy method where GGG substrates are dipped into the flux. 
The growth parameters are controlled to obtain films with in-plane magnetization
and virtually no domain activity, which makes them ideally suited for 
magnetooptic imaging. The Faraday rotation spectra were measured across the visible
range of wavelengths. To interprete the spectra we 
have presented a simple model based on the existence of two optical 
transitions of diamagnetic character, one tetrahedral and one octahedral. 
We find excellent agreement with our experimental results within the 
visible range. It is shown that the Faraday rotation changes significantly with the 
amount of substituted gallium and bismuth. Furthermore, the experimental 
results confirm that the magnetooptic response changes linearly with the 
bismuth substitution.
 
It would be of considerable interest to observe the behaviour at low and high 
temperatures, which will be a topic of future investigations.   

\acknowledgements
We thank Dr. Alexander Solovyev for help to grow samples 
1, 2 and 4, and Dr. Muriel Erambert at the Mineral-Geological 
Museum for assisting in the EMP-measurements. This research has 
been financially supported by the Norwegian Research Council 
and Tandberg Data ASA.

\newpage

\begin{figure}
\caption{Setup for measurement of the Faraday spectra. The analyzer is set at 
45 degrees from extinction to maximize the signal. The polarize the light we
use a sheet polarizer, whereas the analyzer is a Glan-Thompson prism.  Together
they have an extinction ratio of $\sim 10^{-4}$. To measure the signal we used an 
ordinary silicon detector with $3\times3$ $mm^{2}$ active area. 
\label{f1}}
\end{figure}

\begin{figure}
\caption{The Faraday rotation for sample 1. The squares are the
experimental data whereas the solid line represents the theoretical fit. The
dashed and dash-dotted lines represent the contribution from the tetrahedral
and octahedral transitions, respectively. 
\label{f2}}
\end{figure}

\begin{figure}
\caption{The Faraday rotation for sample 2 (circles) and
3 (diamonds). The solid and dashed lines represent the theoretical fit 
for sample 2 and 3, respectively. 
\label{f3}}
\end{figure}

\begin{figure}
\caption{The Faraday rotation for sample 4. 
The dashed line represents a theoretical estimate of the Faraday rotation 
when the gallium substitution is identical to that of sample 4, but with a 
bismuth substitution of $x=1.5$. The dash-dotted line is similar to the dashed
line, but with $x=2$, showing a peak value $>9$ $deg/ \mu m$.
\label{f4}}
\end{figure}

\begin{figure}
\caption{The basic molecular-orbital energy-level diagram. 
(+1) and (-1) represents right and left-hand circular polarization, respectively. 
Note that there are two transitions which influences the Faraday rotation; one
associated with the tetrahedral site, the other with the octahedral site.
The transitions are assumed to follow the selection rules for electric dipole transitions. 
\label{f5}}
\end{figure}

\begin{figure}
\caption{The Faraday rotation as a function of bismuth content. The experimental points
have the same notation as in the previous figures, and are represented for two
different energies, 1.97 eV (630nm) and 2.3 eV (540nm). The lines are the
theoretical curves for the same energies. 
\label{f6}}
\end{figure}

\newpage

\begin{tabular}{c c c c c c c c}
\hline \hline
Sample nr.\,\,\,\,\,\,\,\,\, 
& Lu\,\,\,\,\,\,\,\,\,
& Y\,\,\,\,\,\,\,\,\, 
& Bi\,\,\,\,\,\,\,\,\,  
& Pb\,\,\,\,\,\,\,\,\,  
& Fe\,\,\,\,\,\,\,\,\,
& Ga\,\,\,\,\,\,\,\,\, 
& t ($\mu m$) \\
\hline
1\,\,\,\,\,\,\,\,\,
& 1.619\,\,\,\,\,\,\,\,\, 
& 0.625\,\,\,\,\,\,\,\,\,  
& 0.683\,\,\,\,\,\,\,\,\, 
& 0.064\,\,\,\,\,\,\,\,\,  
& 3.842\,\,\,\,\,\,\,\,\,
& 1.144\,\,\,\,\,\,\,\,\,  
& 7.5   
\\

2\,\,\,\,\,\,\,\,\,
& 2.141\,\,\,\,\,\,\,\,\,  
& 0\,\,\,\,\,\,\,\,\,  
& 0.796\,\,\,\,\,\,\,\,\, 
& 0.065\,\,\,\,\,\,\,\,\,  
& 3.887\,\,\,\,\,\,\,\,\,  
& 1.088\,\,\,\,\,\,\,\,\,  
& 4.0
\\

3\,\,\,\,\,\,\,\,\, 
& 2.166\,\,\,\,\,\,\,\,\,  
& 0\,\,\,\,\,\,\,\,\,  
& 0.874\,\,\,\,\,\,\,\,\, 
& 0.037\,\,\,\,\,\,\,\,\,  
& 3.937\,\,\,\,\,\,\,\,\, 
& 0.969\,\,\,\,\,\,\,\,\, 
& 4.0

\\
4\,\,\,\,\,\,\,\,\, 
& 2.186\,\,\,\,\,\,\,\,\,  
& 0\,\,\,\,\,\,\,\,\,  
& 0.818\,\,\,\,\,\,\,\,\, 
& 0.062\,\,\,\,\,\,\,\,\,  
& 3.744\,\,\,\,\,\,\,\,\, 
& 1.170\,\,\,\,\,\,\,\,\, 
& 2.6 \\
\hline \hline
\end{tabular} \\ 
\\
\\
\\
$\bf{Table}$ $\bf{1}$: The thickness and chemical composition of the four 
samples.

\newpage

\begin{tabular}{c c c c c c c c}
\hline \hline
site\,\,\,\, 
&                \,\,\,\,
&                \,\,\,\,
& $Nf_{i}/x$     \,\,\,\,
&                \,\,\,\,\,\,\,\,\,\,\,\,\,\,\,\,
& $\Delta_{i}$ \,\,\,\,
& $\omega_{i}$ \,\,\,\,  
& $\Gamma_{i}$ \\
\hline
\,\,\,\,
& Sample 1\,\,\,\, 
& Sample 2\,\,\,\,  
& Sample 3\,\,\,\,
& Sample 4\,\,\,\,\,\,\,\,\,\,\,\,\,\,\,\,
& \,\,\,\,
& \,\,\,\,
&  \\
a\,\,\,\,
& $1.72*10^{23}$\,\,\,\, 
& $1.76*10^{23}$\,\,\,\,  
& $1.88*10^{23}$\,\,\,\,
& $1.69*10^{23}$\,\,\,\,\,\,\,\,\,\,\,\,\,\,\,\,
& 0.27\,\,\,\,
& 3.15\,\,\,\,
& 0.47  \\
d\,\,\,\,
&$-3.90*10^{22}$\,\,\,\,  
&$-4.00*10^{22}$\,\,\,\,
&$-4.26*10^{22}$ \,\,\,\,
&$-3.85*10^{22}$\,\,\,\,\,\,\,\,\,\,\,\,\,\,\,\,\,
& 0.11\,\,\,\,  
& 2.51\,\,\,\,
& 0.38  \\
\hline \hline
\end{tabular} \\
\\
\\
\\
$\bf{Table}$ $\bf{2}$: The parameters found to give the best fit between  
eq. (\ref{a4}) and the experimental data. $Nf_{i}/x$ is measured in 
$[cm^{-3}]$, whereas $\Delta_{i}$, $\omega_{i}$ and $\Gamma_{i}$ are given 
in $[eV]$. Note that the tetrahedral and octahedral sites are given different
signs, since they contribute oppositely to the Faraday rotation.

\newpage
\centerline{\includegraphics[width=14cm]{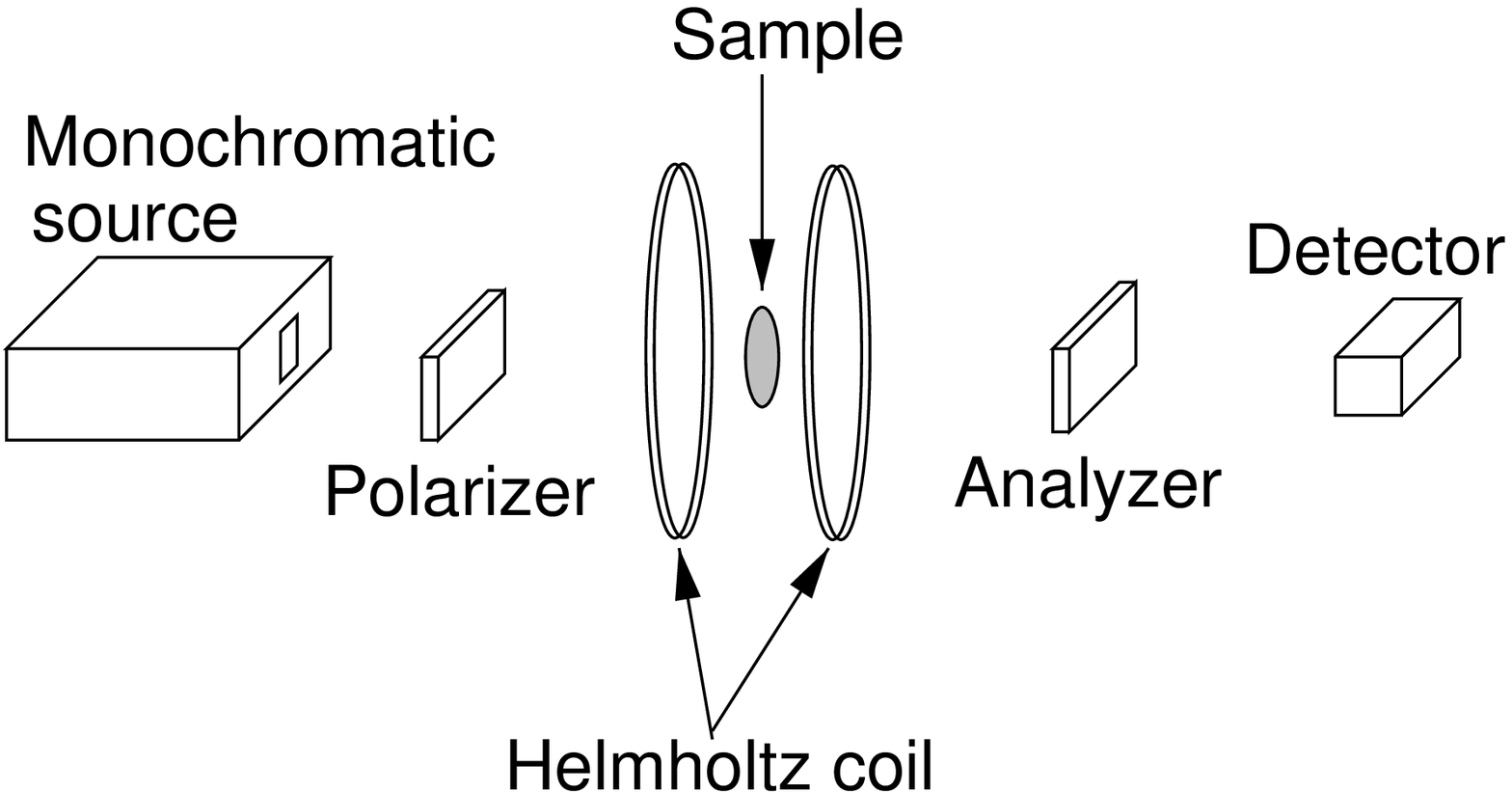}}
\vspace{2cm}
\centerline{Figure~\ref{f1}}

\newpage
\centerline{\includegraphics[width=14cm]{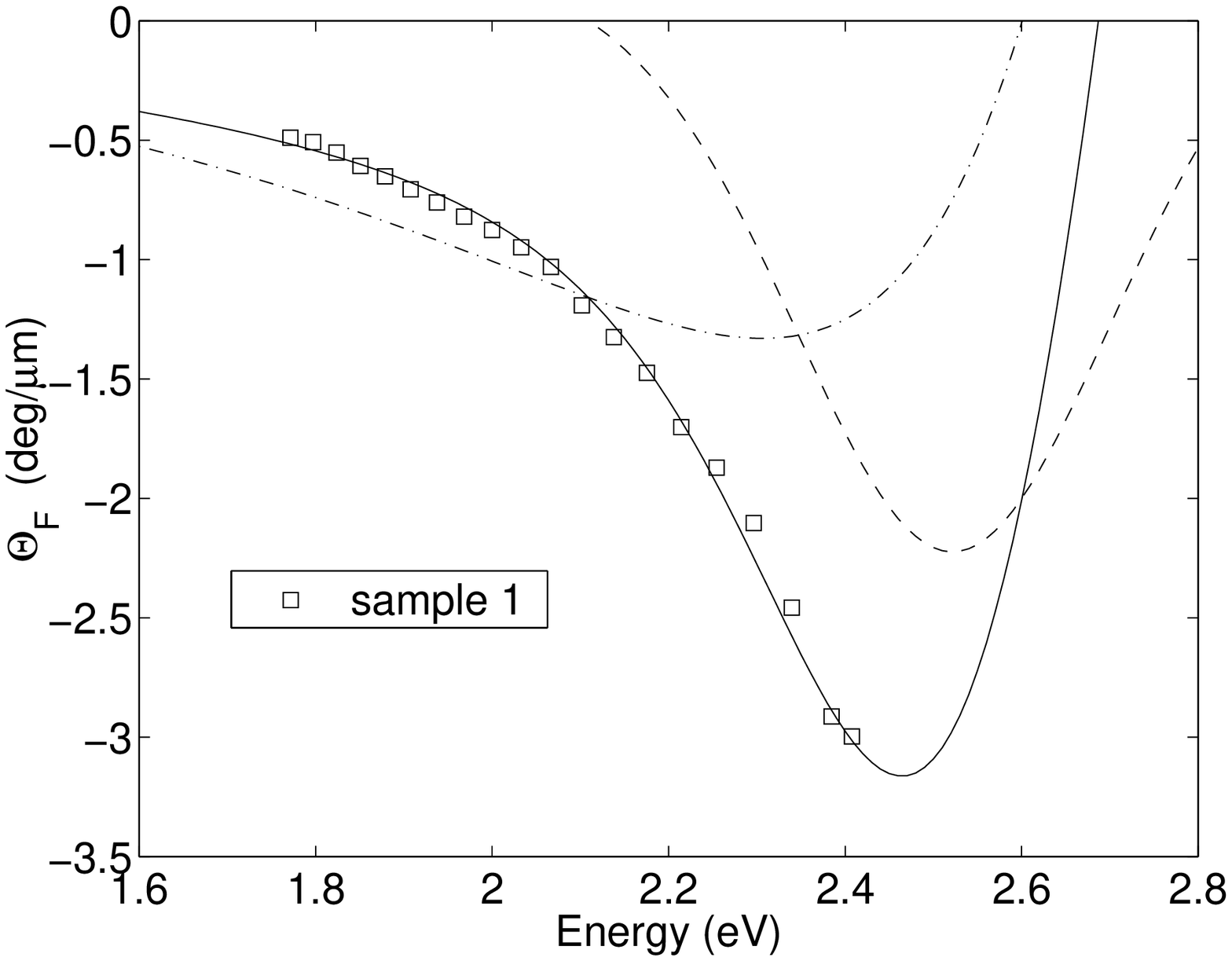}}
\vspace{2cm}
\centerline{Figure~\ref{f2}}

\newpage
\centerline{\includegraphics[width=14cm]{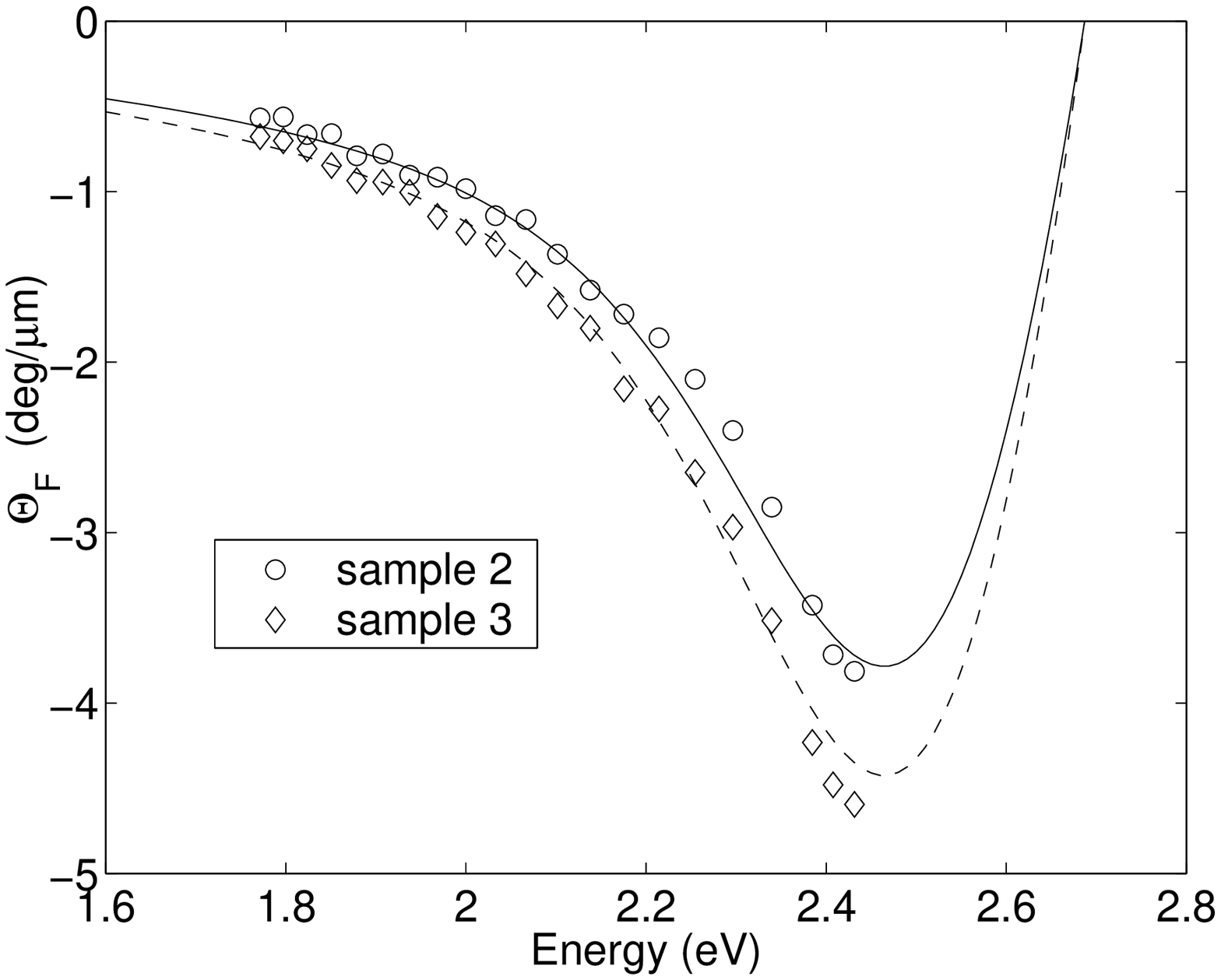}}
\vspace{2cm}
\centerline{Figure~\ref{f3}}

\newpage
\centerline{\includegraphics[width=14cm]{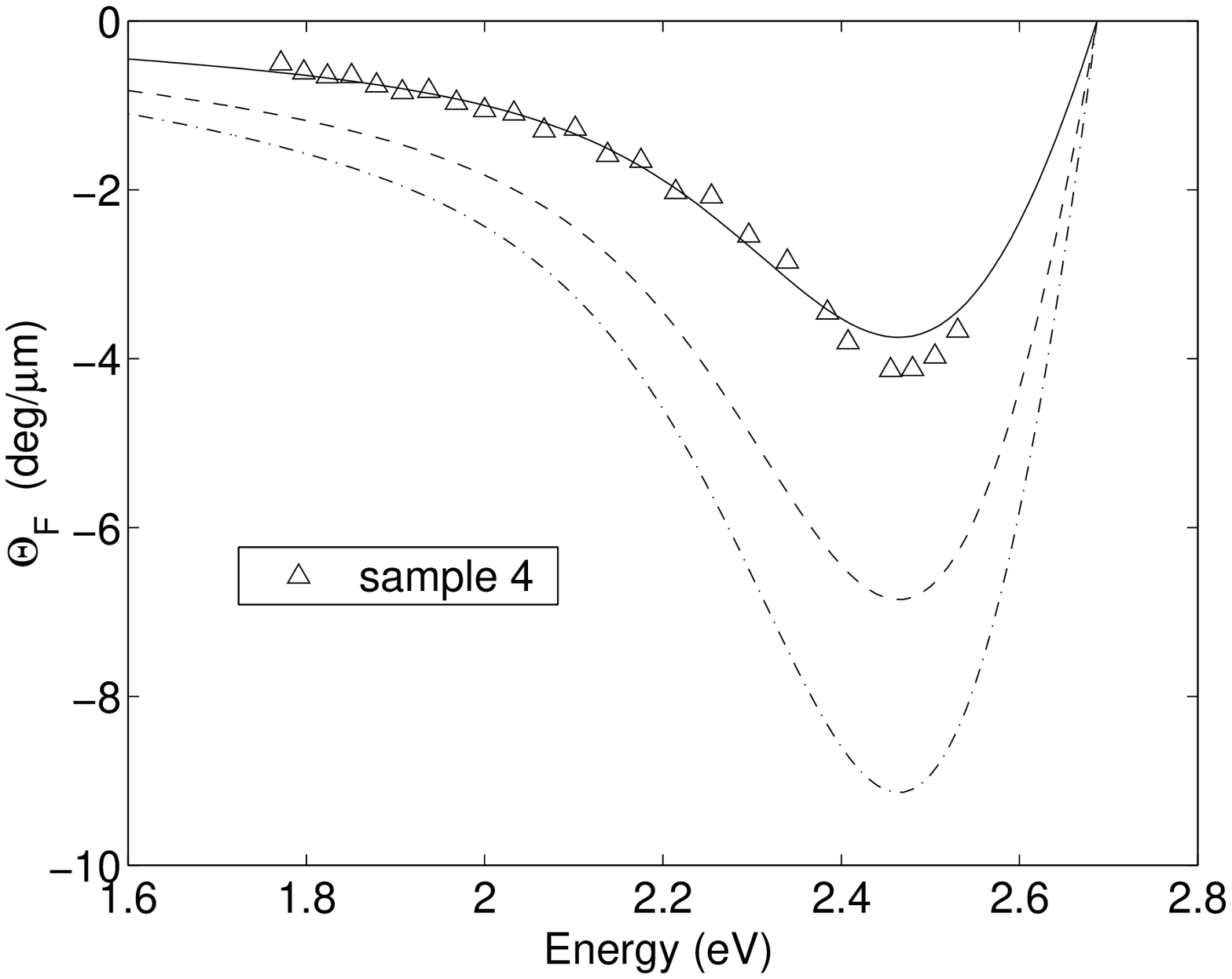}}
\vspace{2cm}
\centerline{Figure~\ref{f4}}

\newpage
\centerline{\includegraphics[width=14cm]{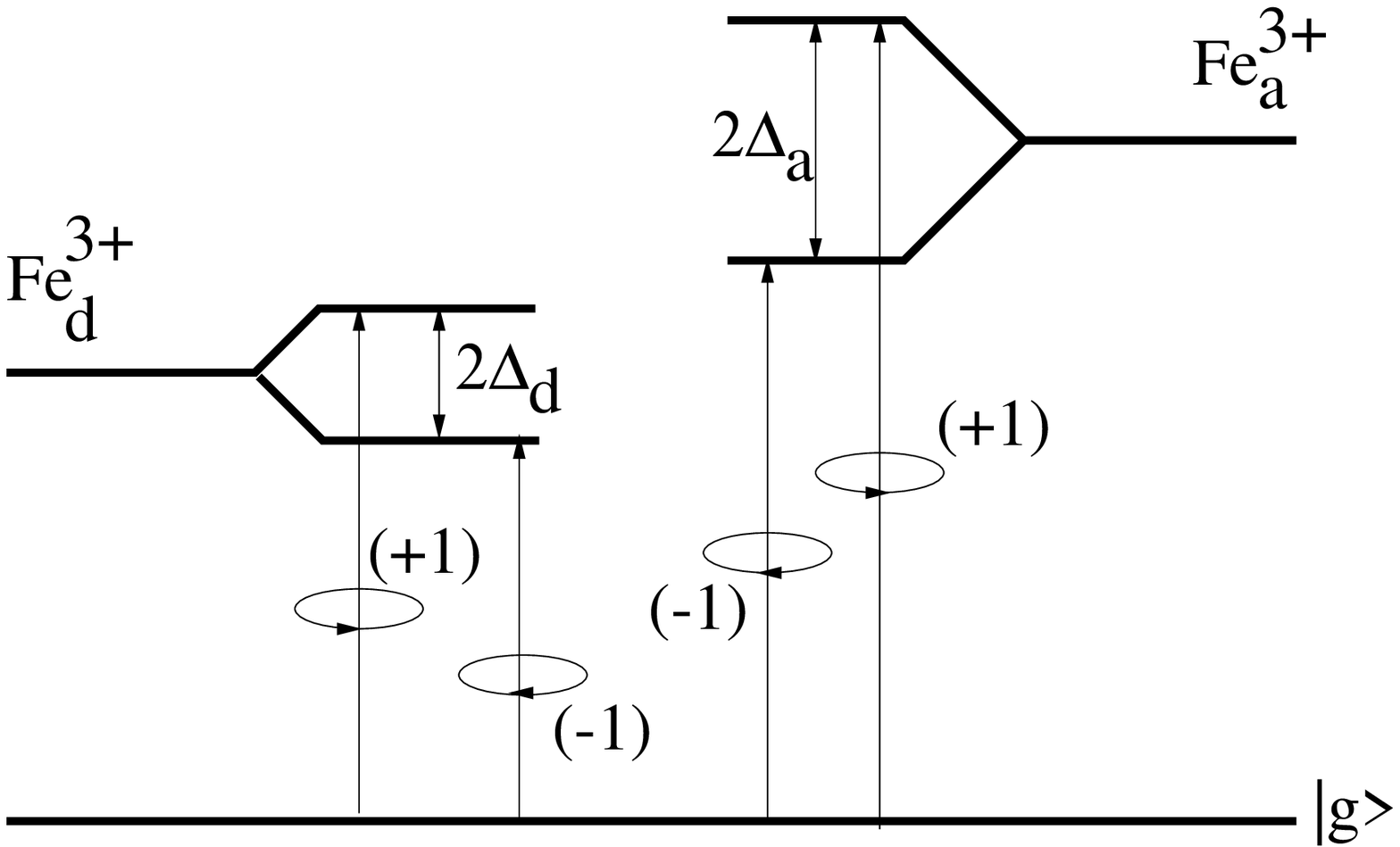}}
\vspace{2cm}
\centerline{Figure~\ref{f5}}

\newpage
\centerline{\includegraphics[width=14cm]{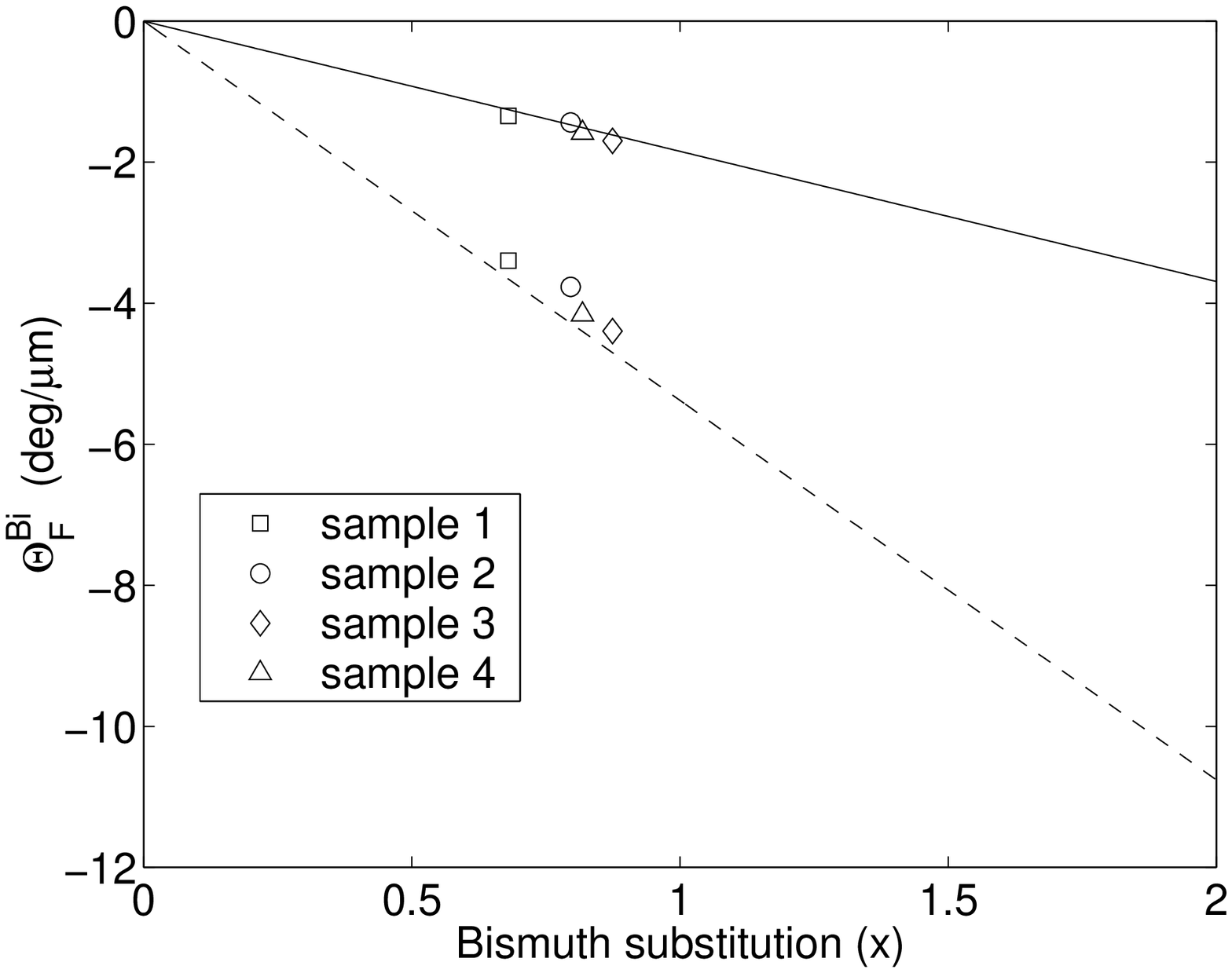}}
\vspace{2cm}
\centerline{Figure~\ref{f6}}

\end{document}